\documentclass[11pt]{article}
\usepackage{latexsym}
\usepackage{graphicx}

\begin{document}

\title{Physics of Econophysics}
\author{Yougui Wang$^1$, Jinshan Wu$^2$, Zengru Di$^1${\footnote{Email: zdi@bnu.edu.cn}}
\\1. Department of System Science at School of Management, \\Beijing
Normal University, Beijing, 100875, P.R.China
\\2. Department of Physics, Simon Fraser University, Burnaby, B.C.
Canada, V5A 1S6}

\maketitle

\begin{abstract}
Econophysics is a new area developed recently by the cooperation
between economists, mathematicians and physicists. It's not a tool
to predict future prices of stocks and exchange rates. It applies
idea, method and models in Statistical Physics and Complexity to
analyze data from economical phenomena. In this paper, three
examples from three active main topics in Econophysics are
presented first. Then using these examples, we analyze the role of
Physics in Econophysics. Some comments and emphasis on Physics of
Econophysics are included. New idea of network analysis for
economy systems is proposed, while the actual analysis is still in
progress.
\end{abstract}

\section{Introduction}
Econophysics is a developing field in recent years. It's a subject
applying and proposing idea, method and models in Statistical
Physics and Complexity into analyzing data coming from economical
phenomena. Economics is a subject about human behavior related
with the management of the resources, finances, income, the
production and consumption of goods and services. So Economics is
usually regraded as a social science. But in some ways, the laws
in Economics are similar with natural science. Although it has to
deal with incentive and human decision, but sometimes the
collective behavior can be described by determinant process, at
least in a statistical way. So the aim of Econophysics is to apply
the idea of natural science as far as well into economics. Maybe
this will disentangle natural laws and human behaviors in
economical phenomena, and end with a new Economics.

Also because of the plenty data records of different systems in
our economy behavior, it's a treasure to physicists, especially to
the one being interested in Complex Systems, in which many
subsystems and many variables interact together. And the
development of Economics also provide many open questions, like
stock price, exchange rate and risk management, which may require
technics dealing with mass data and complex systems.

Physics tries to construct a picture of the movement of the whole
nature. Mechanism is the first topic cared by physicists. So
trying to describe and understand the phenomena is the first step
of econophysicists facing the mass data in economical phenomena.
Till now, we have to say, the most works in Econophysics are
empirical study of different phenomena to discover some universal
or special laws, and also some initial effort about models and
mechanism.

Therefor, in this talk, we will begin with three examples of
empirical works in Econophysics, and discuss very shortly about
the corresponding models and mechanism. Focus will be on the
Physics of Econophysics, to present the power of Physics to
Econophysics and some benefit which Physics will get from
Econophysics.

\section{Three main topics of Econophysics}
Recent works in Econophysics mainly in three objects. First one is
the time series of stock prices, exchange rates and prices of
goods. Size of firms, GDP, individual wealth and income are the
second topic, which can be regarded as wealth of different
communities. The third one is network analysis of economical
phenomena.

\subsection{Fluctuation of stock prices and exchange rates}
The prices of stocks are recorded every minute or every few second
everyday in stock market all over the world. The price of a stock
is driven by many factors, such as the whole economy environment,
achievement of enterprise, the prices of other stocks, and by the
buying or selling activity of stockholders. At the same time, the
behavior of stockholders is effected by the price, and further
more, everyone has his/her own decision which is different with
each other, but effected each other. So such phenomena seem
complex. While every enterprise has its own characters, and every
stockholder decide his/her behavior on his/her own knowledge,
information and belief, and every stock market has its own
environment, the empirical study shows some common stylized facts
valid for almost all stocks.

\begin{figure}[tbp]
\centering \includegraphics[width=4in]{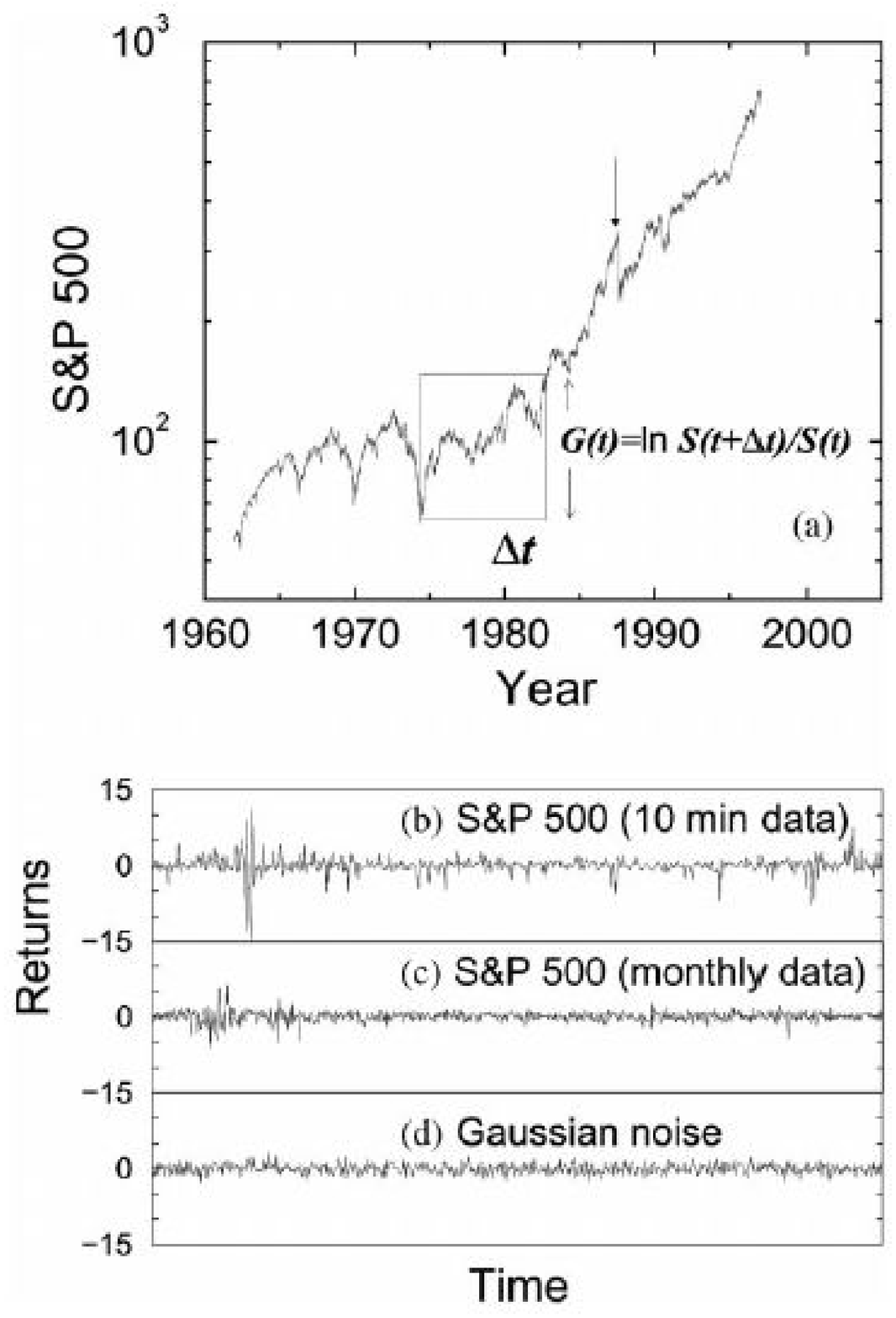}
\caption{Typical time series of stock price and return}{extracted from
\cite{correlation}, time series of stock price. The first two
figures at bottom are time series of return while the last one is
a Gaussian noise. } \label{figdata}
\end{figure}

A typical time series of stock price, S$\&$P500 index, denoted as
$S\left(t\right)$, is showed in figure \ref{figdata}. Actually
S$\&$P500 is a stock index, which is a weighted mean value of
stocks in a market, can be used as a indicator of stock price.
Some papers use the data of indexes, some use individual stock,
and also some paper investigate all stocks in a market as an
ensemble of stocks. In this talk, we just use analysis of
individual stock as examples.

Because economy is in growth, so the time series of stock price
has a long term trend to increase. This means it's nonstationary.
So other than the original price, other quantities like different
and return may be better to use as analysis object. The difference
is defined as
\begin{equation}
D_{\Delta t}\left(t\right)=S\left(t+\Delta t
\right)-S\left(t\right),
\end{equation}
in which $\Delta t$ is the time step to sample the time series. It
can be the time step of record, or a large time scale. Return is
defined as
\begin{equation}
G_{\Delta t}\left(t\right)=\ln\left(S\left(t+\Delta t
\right)\right)-\ln\left(S\left(t\right)\right).
\end{equation}
It's equivalent with $\frac{D_{\Delta
t}\left(t\right)}{S\left(t\right)}$ when $\Delta t$ is small
enough.

Most works use return as object time series. The figures in the
lower part of figure \ref{figdata} show examples of
$G\left(t\right)$, while the last one is a Gaussian noise signal
for comparison.

A statistical analysis of one time series can be classified as two
parts, the distribution properties which dismiss the time
information, and the autocorrelation analysis which mainly takes
time into account.

\subsubsection{Distribution properties}

\begin{figure}[tbp]
\centering
\includegraphics[width=4in]{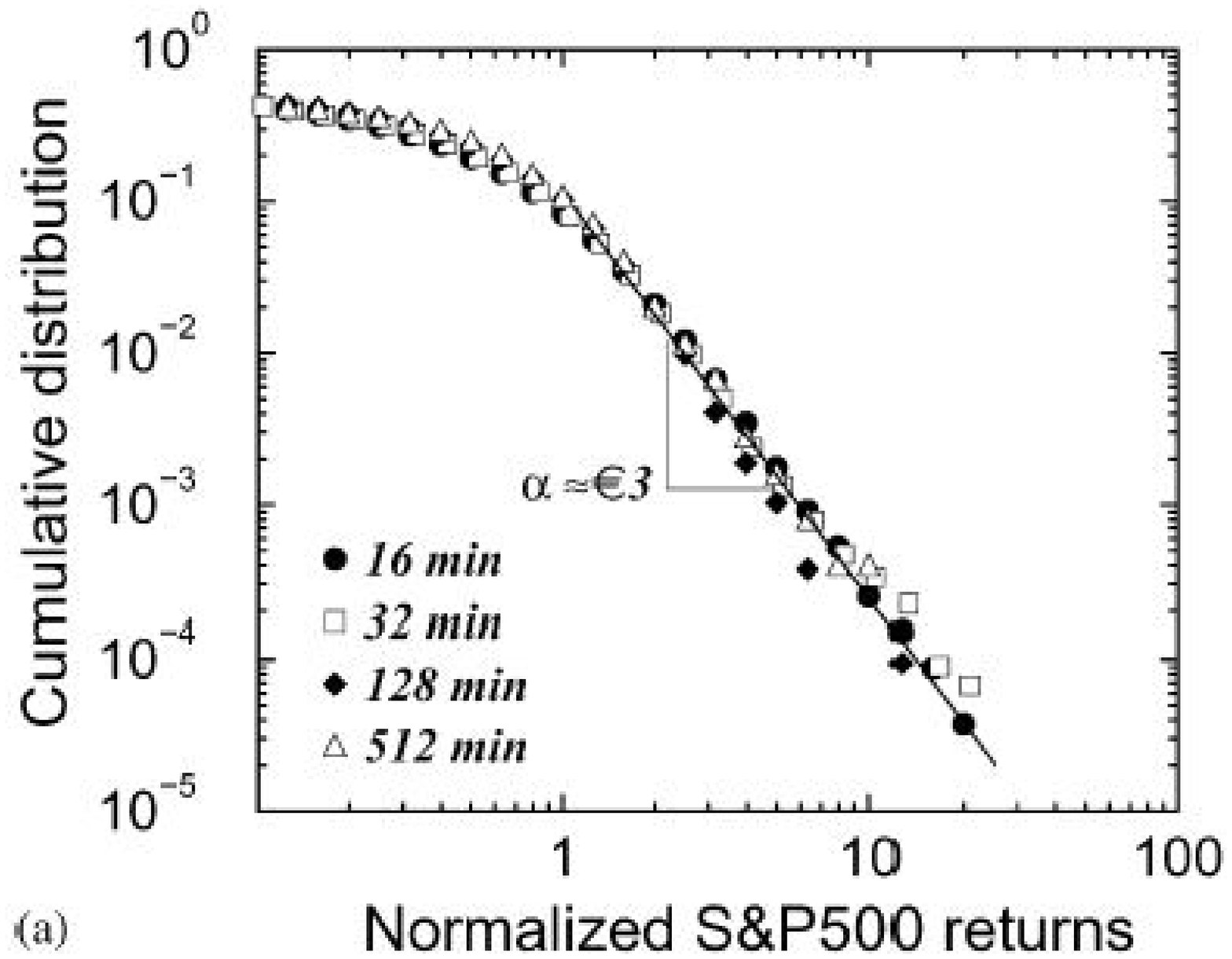}
\includegraphics[width=4in]{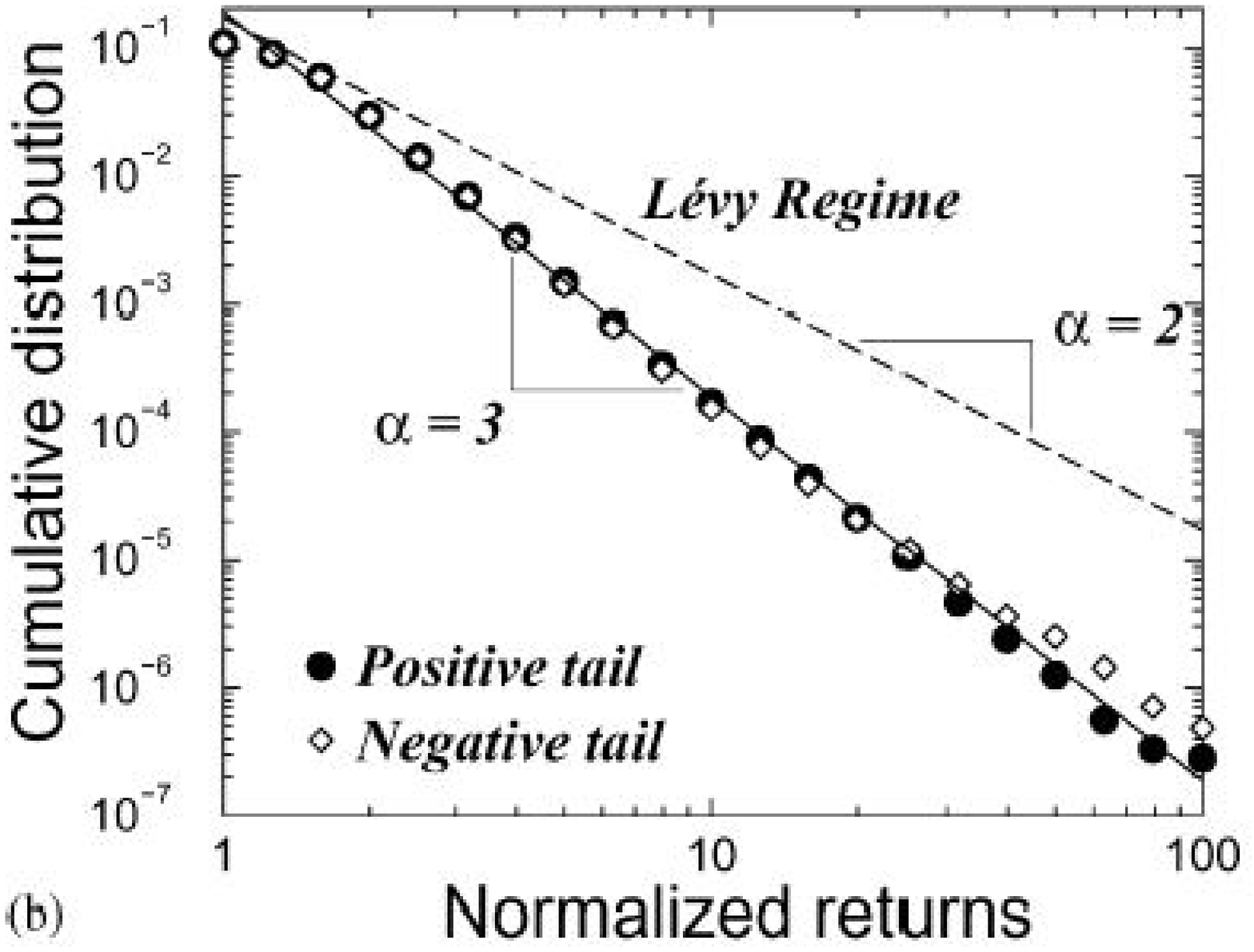}
\caption{distribution of return}{extracted from
\cite{correlation}, Log-normal for central part and power law
heavy tail. } \label{figreturn}
\end{figure}

\begin{figure}[tbp]
\centering
\includegraphics[width=4in]{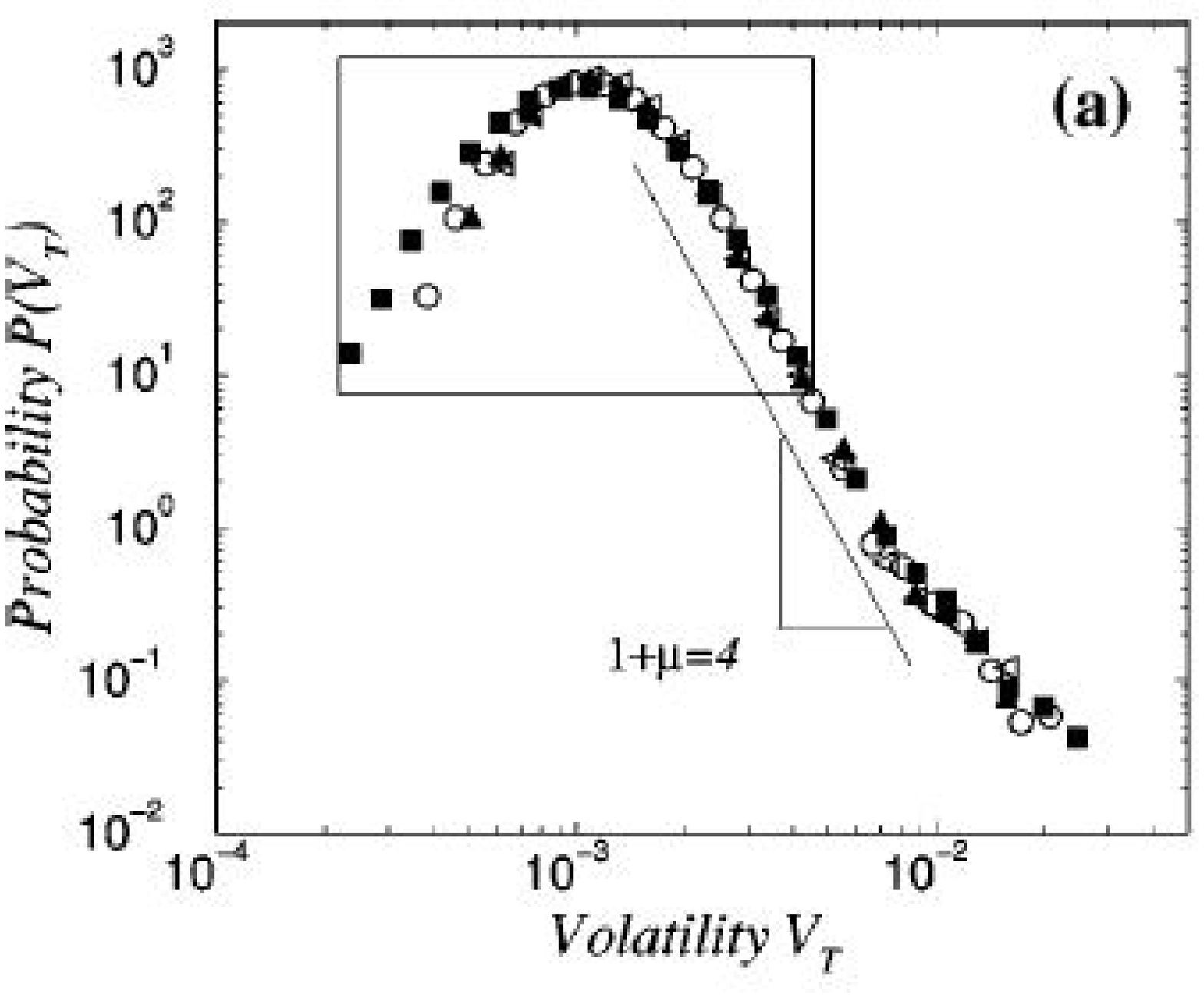}
\includegraphics[width=4in]{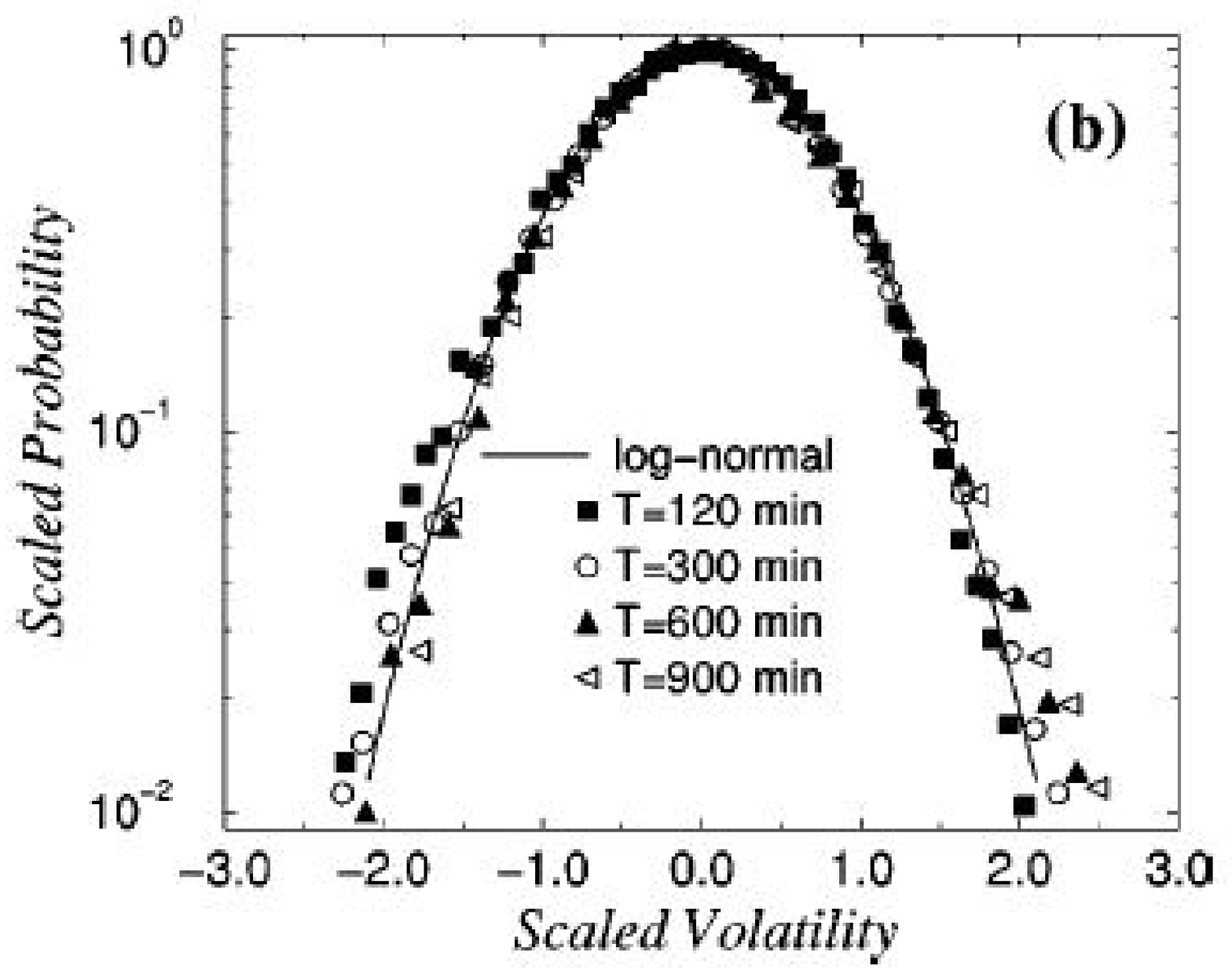}
\caption{distribution of volatility}{extracted from
\cite{volatility}, Log-normal for central part and power law
heavy tail. } \label{figvola}
\end{figure}

The frequency account of a data set formed by collecting all the
return values will give us the distribution, as shown in figure
\ref{figreturn}. Detailed fitting shows the central part is a
log-normal distribution ($p(x)\sim e^{-\ln^{2}x}$) while the tail
is a power law distribution ($p(x)\sim x^{-\alpha}$). The more
important thing here is the universality. The distribution shape
is independent on the time scale ($\Delta t$), and is a common
distribution for different stocks in different markets, even in
different countries. When an empirical statistical result is
universal, we have to ask for the common nature behind it.
Another distribution properties is about the volatility of stock,
which is related with risk. So its characters is important for
risk management. Usually it's defined by local variation,
\begin{equation}
V_{T}\left(t\right)=\sum^{t+T}_{\tau=t}\left(G\left(\tau\right)-\bar{G}_{T}\right)^2
\end{equation}
in which $T=n\Delta t$ is a time window moving along with the
time, and $\bar{G}_{T}$ is the mean value of $G\left(t\right)$ in
the window, as
\begin{equation}
\bar{G}_{T}=\frac{1}{n}\sum^{t+T}_{\tau=t}G\left(\tau\right).
\end{equation}
In some papers, volatility is also defined as
\begin{equation}
V_{T}\left(t\right)=\sum^{t+T}_{\tau=t}\left|G\left(\tau\right)\right|,
\end{equation}
in which absolute value is equivalent with square, and we don't
care about the mean value of $V_{T}$, which can be set to be zero
when we analysis the distribution function or autocorrelation.
Also it was found that the distribution function is universal for
different stocks in different market during different time.
Similarly the center part is log-normal, while the tail is power
law, which is shown in figure \ref{figvola}.

\subsubsection{Autocorrelation}

Besides the distribution property, most information of time series
is included in time. Now we present the result of autocorrelation
analysis of return and volatility\cite{correlation}. The
autocorrelation of a stationary time series is defined as
\begin{equation}
C\left(\tau\right) = \frac{\langle
G\left(t+\tau\right)G\left(t\right)\rangle-\langle
G\left(t+\tau\right)\rangle\langle G\left(t\right)\rangle}{\langle
G^{2}\left(t\right)\rangle-\langle G\left(t\right)\rangle^{2}}
\end{equation}
It can be investigated by spectrum analysis. But for a
nonstationary one, a recently developed detrend fluctuation
analysis (DFA)\cite{dfa} is commonly used.

\begin{figure}[tbp]
\centering
\includegraphics[width=4in]{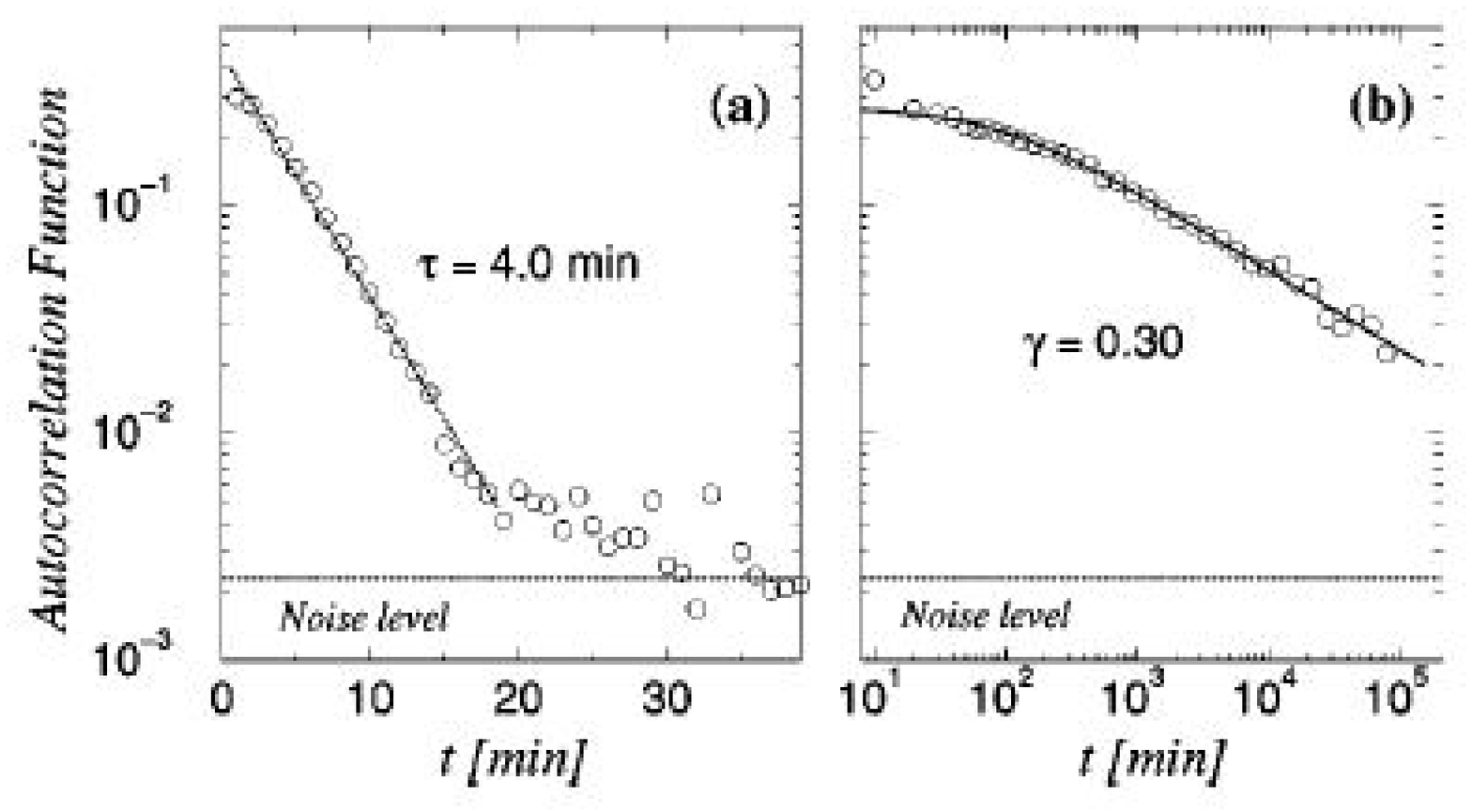}
\caption{Autocorrelation of return and volatility}{extracted from
\cite{volatility}, Exponential decay for return while power law decay for volatility}
\label{figauto}
\end{figure}

In figure \ref{figauto}, the autocorrelation functions of return and
volatility are plotted together. We can find an exponential drop
off in return with a time scale of minute, while a power law
decrease in volatility without a finite time scale. Think about
this phenomenon, a time series almost without an autocorrelation,
but a extremely high autocorrelation in absolute value, or local
variation. It's amazing. The fast dropping off guarantee the
validness of Efficient Market Hypothesis, while the long time
autocorrelation in volatility make it possible to construct a
theory of risk management. So such works will boost the
development of risk management, even a reformation.

\subsubsection{Price and volume}
\begin{figure}[tbp]
\centering
\includegraphics[width=5in]{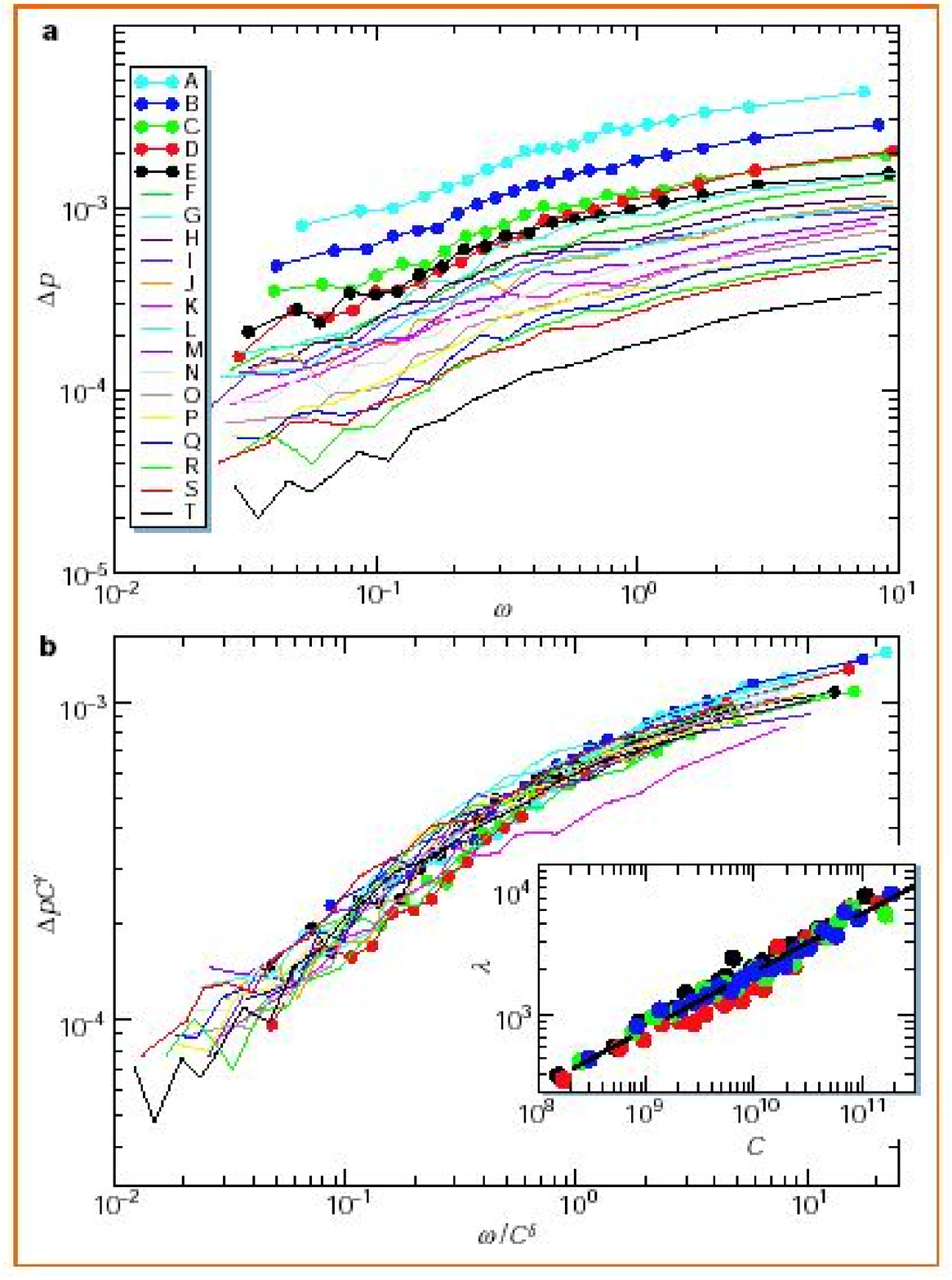}
\caption{Master Curve for the impaction from volume to price}{extracted from
\cite{mastercurve}, data collapse onto a master curve}
\label{figdemand}
\end{figure}

All the analysis above looks like kinetics, which solve the
question how to describe the movement and what's the movement. The
next question in the tradition of Physics is how can such movement
happen. So next step, let's think about what are the factors
effect the stock price. And again, we may try to keep our eyes on
empirical study as far as we can. Demand and supply decide the
price is a central law in Economics. Although we know it's for
price of goods, maybe it will still be valid for stocks. So it
leads us to empirical study of order book of stocks, which study
the relation between difference of prices and the transaction
volume\cite{mastercurve,demand}. For a individual stock, they
recorded transaction volume $\omega$ as the total volume of every
transaction before the price changed, and define the difference of
the logarithm of the price now and price before such change as
price shift,
\begin{equation}
\Delta p\left(t_{i+1}\right)=\ln S\left(t_{i+1}\right) - \ln
S\left(t_{i}\right).
\end{equation}
Then plots of price shift ($\Delta p\left(t\right)$) vs
transaction volume ($\omega\left(t\right)$) are presented in the
up part of figure \ref{figdemand}. In the lower part, the authors found all
curves can be collapse onto a common line by rescale. So it's also
a universal law for all stocks.

From the above results, it seems that stock price is only
determined by transaction volume. But it's sad to say, the
transaction volume is also decided by price. No direct way to
predict transaction volume. It should be decided together with
price by other predictable or known variables. So let's say if we
have only one stock, and the whole history of this stock is
already known, the achievement and activity of the enterprise is
also predictable by other ways, and so is the external economy
environment, at least in a statistical way, which means if they
are random variables we know the distribution and correlation, in
such condition, is the future of this stock determinant, and is it
predictable or chaotic? Or at least we can reproduce the similar
data with the same statistical characters as the empirical data?
If it's possible, what's the central variables, and how it can be
generalized into a stock ensemble, not only one stock?

\subsubsection{Toy models}

The questions above ask for a mechanism model of stocks. Maybe
it's not very possible to reproduce the exact time series, but if
the stylized statistical facts are reproduced, the model is well
done in physical sense, although not in a sense of making money.
Let's check what's the central variable left after so many things
settled down by us including initial condition (even history),
boundary condition (only one stock), external variables
(enterprise and environment). So the only one thing left here is
how do stockholders buy and sell the stock under different price
and how does the price effected by the transaction.

The first idea here is activities of all stockholder are effected
each other. Such interaction maybe is indirectly through the price
and market, or by external way such as personal relationship. As a
tradition in Physics, a first approximation is treat every
stockholder independently, so they will only effected each other
through market. Like in spin model, every stockholder will has a
unit volume can buy or sell every time. Buying will improve the
price while selling lowers the price. Everyone is trying to make
more money in this game. So till now, a toy model has been
constructed for mechanism of stock price. When the detail of
benefit evaluation of every player and the effect on price by one
unit volume is set, this toy model will evolute in its own way, of
course when some specific behavior of all external variables are
also settled down.

Although it's only a toy model, we also can test some fundamental
knowledge, such as benefit and rational agent, and also try
different form of external variables. For example, we can take for
granted that external variables are random signals with fixed
distribution and without autocorrelation of any order. So our task
will be how can we construct our model to reproduce the
autocorrelation behavior in empirical study from no
autocorrelation input external data.

And then, if the output data is totally incomparable, maybe we
have to add something we dismissed, like the relation between
stocks. You know a phone call from your close friend may change
your decision. So it's very possible we have to take such
interaction into consideration. The model in \cite{spin}, is a
representative one of such toy models. Although many different
interaction forms we can try, or even we can coevolute the
interaction strength together with the stock price, it's possible
that the output data is still incomparable with empirical one.
Then, we will have to include the interaction between different
stocks, and maybe further more a coevolution system including the
behavior of enterprises.

Oh, no, wait a minute, this is not on the way of physics now. More
and more variables, more and more subsystems, uglier and uglier
picture. It shouldn't. The Physics of Complex System tells us
maybe only a few ones rule the system. So the toy model maybe
imply something valuable. Now we come back to empirical study and
toy model, but in another way, the way keeping Physics in mind.

\subsubsection{Goods, options and others}
Not only the stocks, also exchange rates, goods and options are
under analysis nowadays. However, the universal results for stocks
seems not valid for other goods and options. The empirical study
of land price\cite{land} gives the high skew and heavy tail
distribution of price ($S\left(t\right)$) and power law
distribution for relative price
($r\left(t\right)=\frac{S\left(t+1\right)}{S\left(t\right)}$. And
empirical study of return of options shows unsymmetrical power law
distribution\cite{option}.

And not only the prices, waiting time can also be take into
consideration. In a real stock market, transactions do not always
happen in every minute or every half minute. It's also a random
variable. And the prices change depending on the transactions.
This is the so-called continuant time stochastic process.
Empirical and model analysis just started
up\cite{time1,time2,time3}.

\subsection{Distribution of firm sizes, GDP, personal income and wealth}
Interaction between different communities such as trade,
cooperation and competition, plays important roles in economy. As
a result of such activities, the wealth distribution carries some
valuable information for researchers to investigate the properties
of such interaction. So the second active main topic in
Econophysics is about the size distribution. For a firm, size can
be measured by employee, sales or capitals, while GDP for a
country, income or wealth for a person.

\subsubsection{Distribution of size}
\begin{figure}[tbp]
\centering
\includegraphics[width=3in]{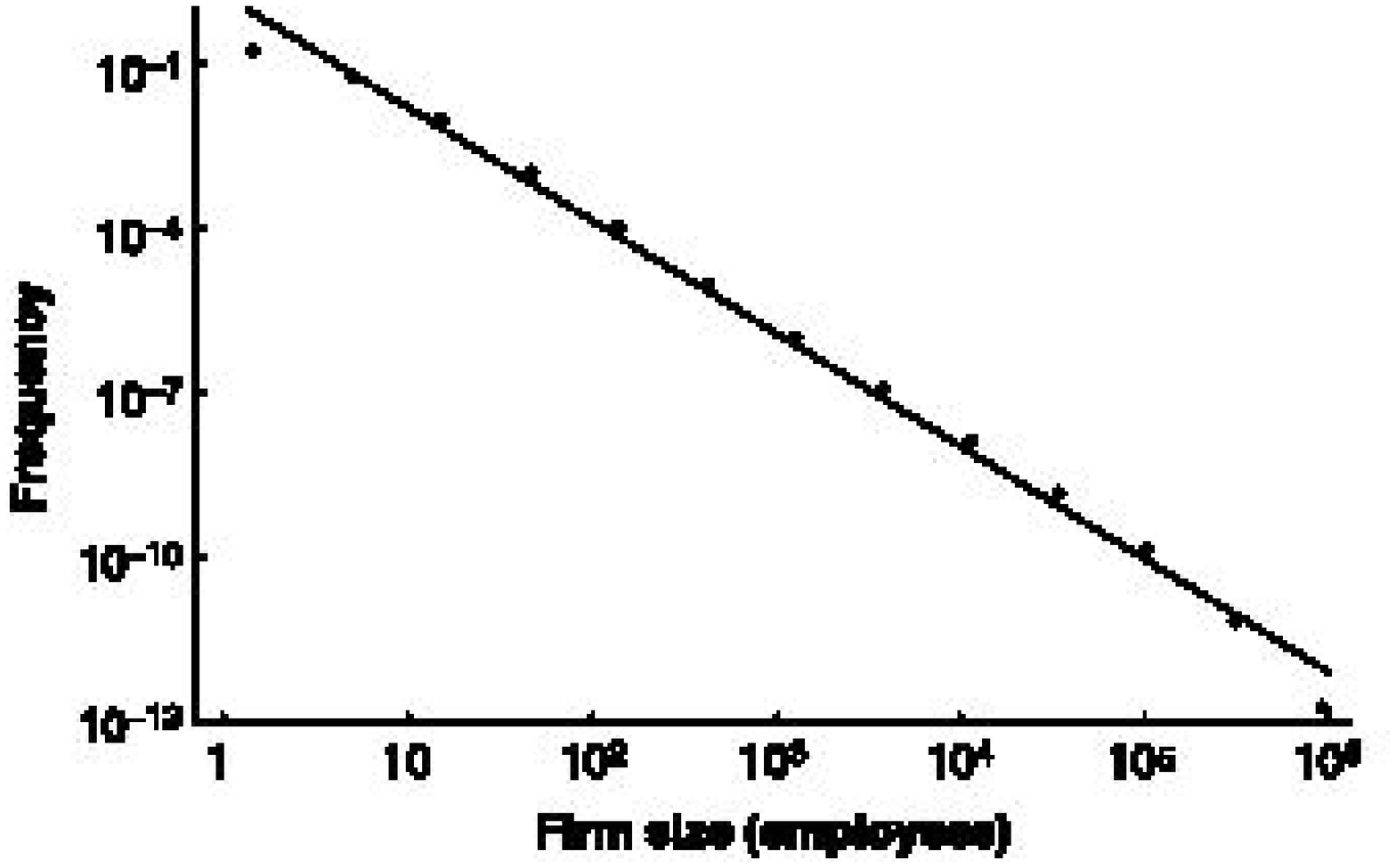}
\includegraphics[width=3in]{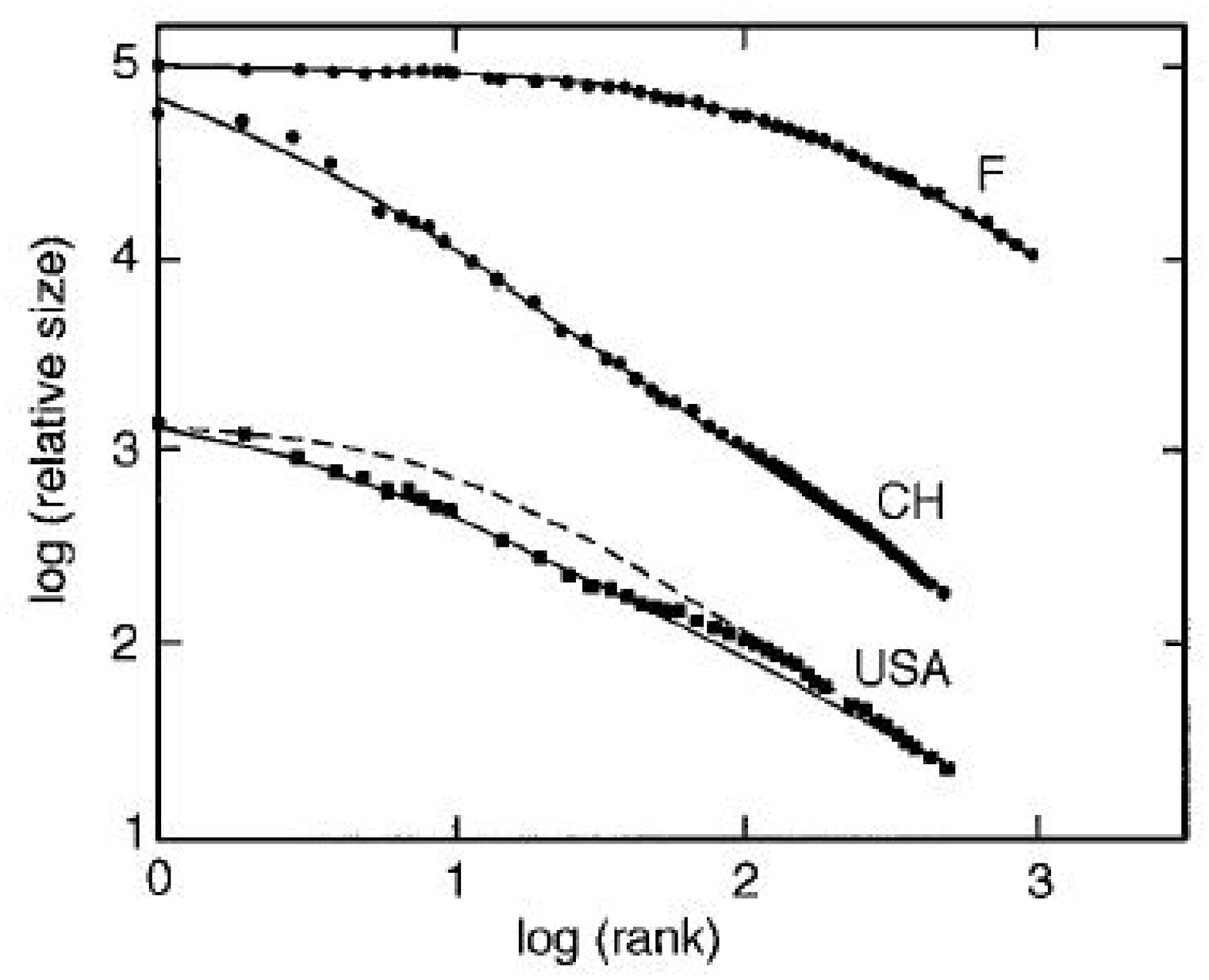}
\caption{Zipf plot of sizes of firms}{extracted from
\cite{firmaxtell, firmsincountries}, universal power law distribution.}
\label{figfirmsize}
\end{figure}

In \cite{firmaxtell}, the author collected data including more
firms especially small firms in a longer history than the database
in \cite{zipf}, so the result of power law distribution seems more
convincing than the log-normal distribution in the later. And the
important character about such distribution is the universality.
Different measurement of size such as total employees, sales,
assets and capitals give the same distribution. And it doesn't
depend on the time, even during the years of significant change of
working force. Further investigation\cite{firmsincountries} shows
it's also a common law in different countries. A typical
distribution is shown in figure \ref{figfirmsize}.

Similar results have been get for GDP of countries all over the
world. Power law distribution of GDP per capita of different
counties has been revealed in \cite{GDP1,GDP2}.

\begin{figure}[tbp]
\centering
\includegraphics[width=3.5in]{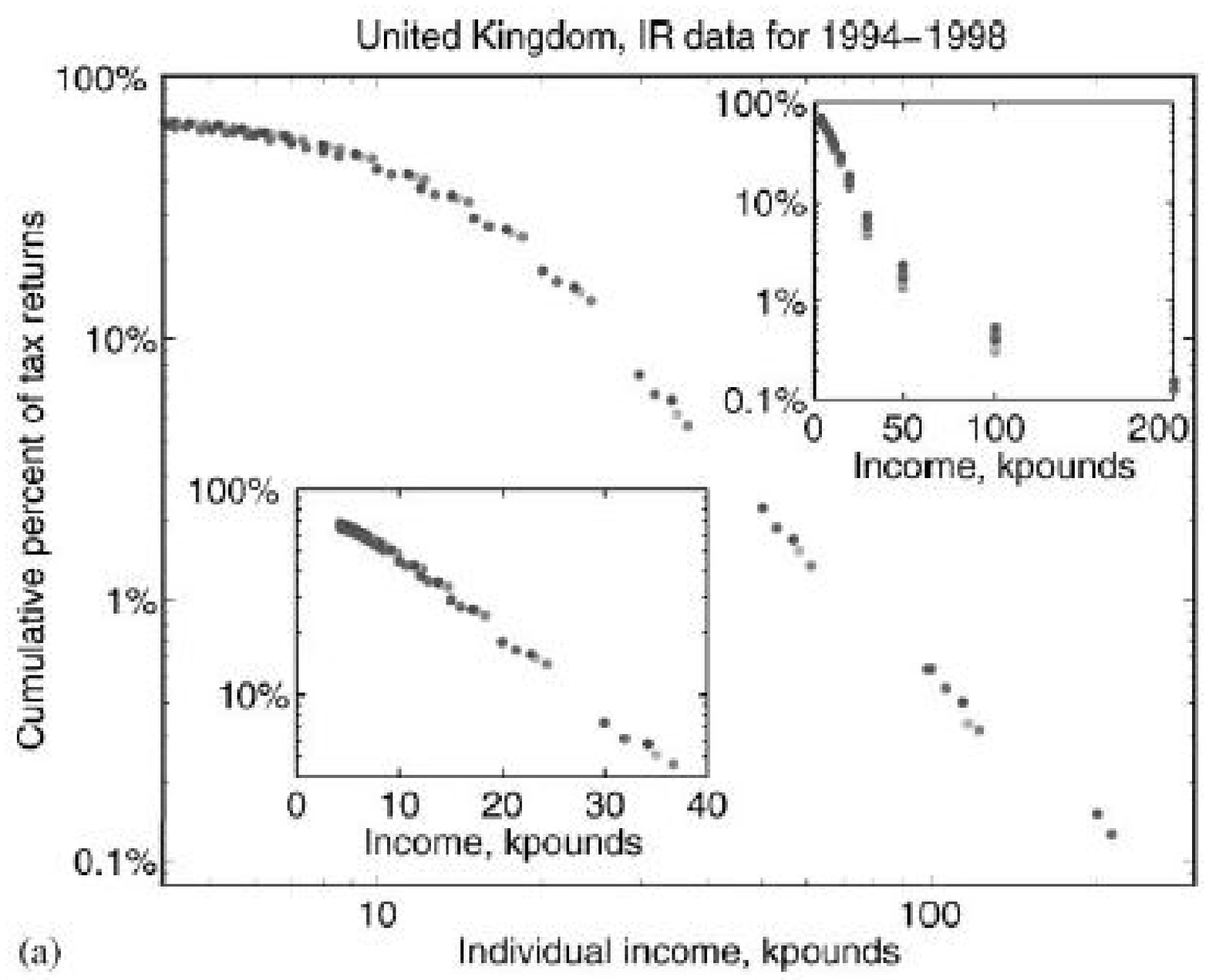}
\includegraphics[width=3.5in]{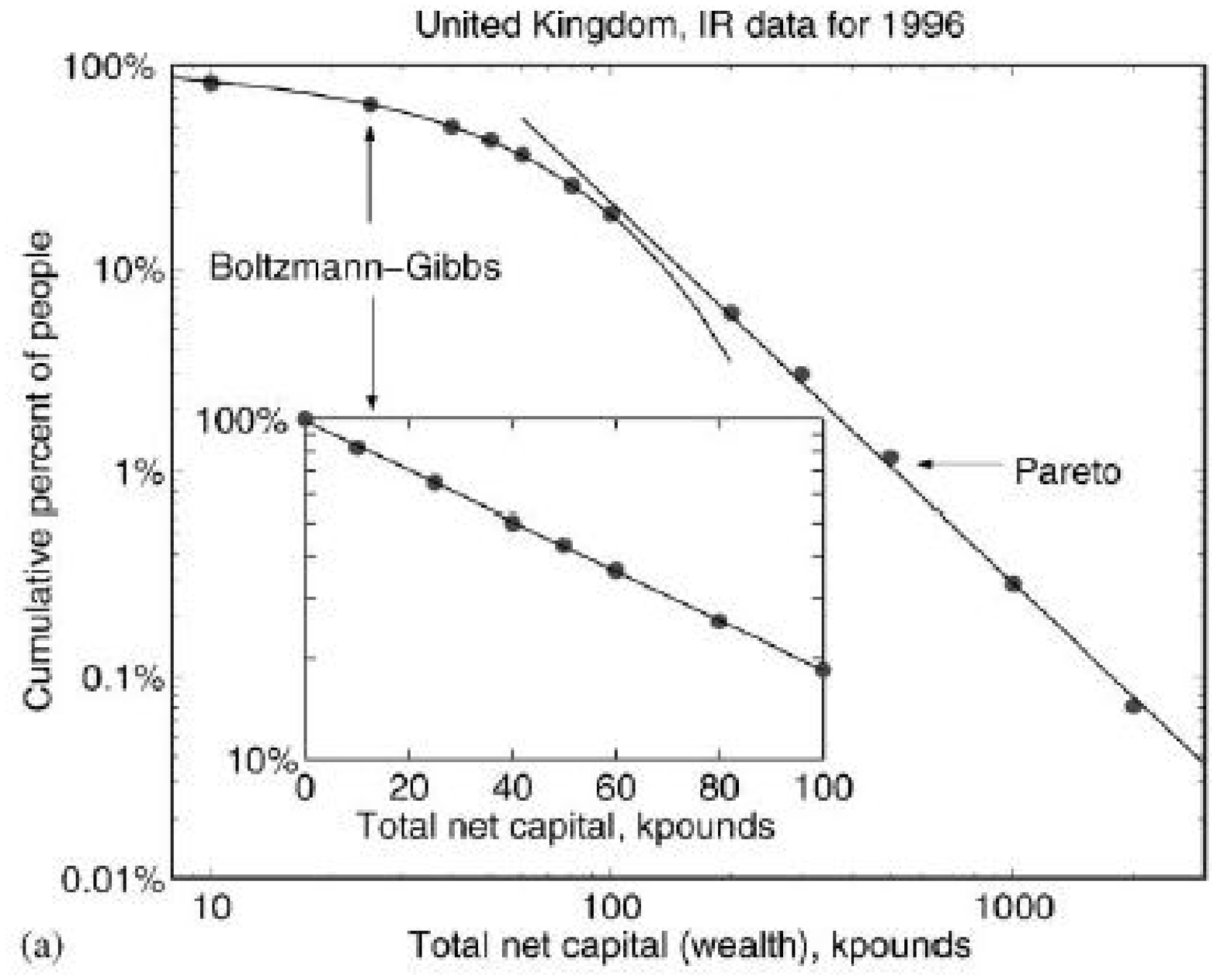}
\caption{Distribution of individual income and wealth}{extracted from
\cite{wealth1}, Exponential distribution for lower part while power law distribution for high tail.}
\label{figwealth}
\end{figure}

For individual such distribution can be analyzed by personal
income or wealth. A typical result\cite{wealth1,wealth2} is shown
in figure \ref{figwealth}. The lower income seems like exponential distribution
while the higher part is power law. From experience of ideal gas,
we know, the equilibrium energy distribution of a random exchange
system is exponential. So maybe in the lower income community, the
cooperation and competition between individuals is in a way
similar with random exchange. But for the higher income part,
different interaction like preferential attachment part more
important role.

\subsubsection{Growth rate}
Growth rate of firm size is defined similarly with return as
\begin{equation}
r\left(t\right)=ln\left(S\left(t+1\right)/S\left(t\right)\right).
\end{equation}
Then in an ensemble of firms, every firm has its own track, and at
every time, we have a cross-section data set including all firms.
In a tradition of Statistical Physics, analysis can be done along
two ways, keeping eyes on individual time series or just dealing
with cross-section data. In an ensemble consist of identical
systems, those two ways will give the same result. However,
although here we can make an assumption that all firms act in a
common way, which is the way we want to find, our ensemble is not
consist of identical systems. So the compromise here was to treat
the firms with the same size as identical systems, and to dismiss
the time information and mix them together.

\begin{figure}[tbp]
\centering
\includegraphics[width=3in]{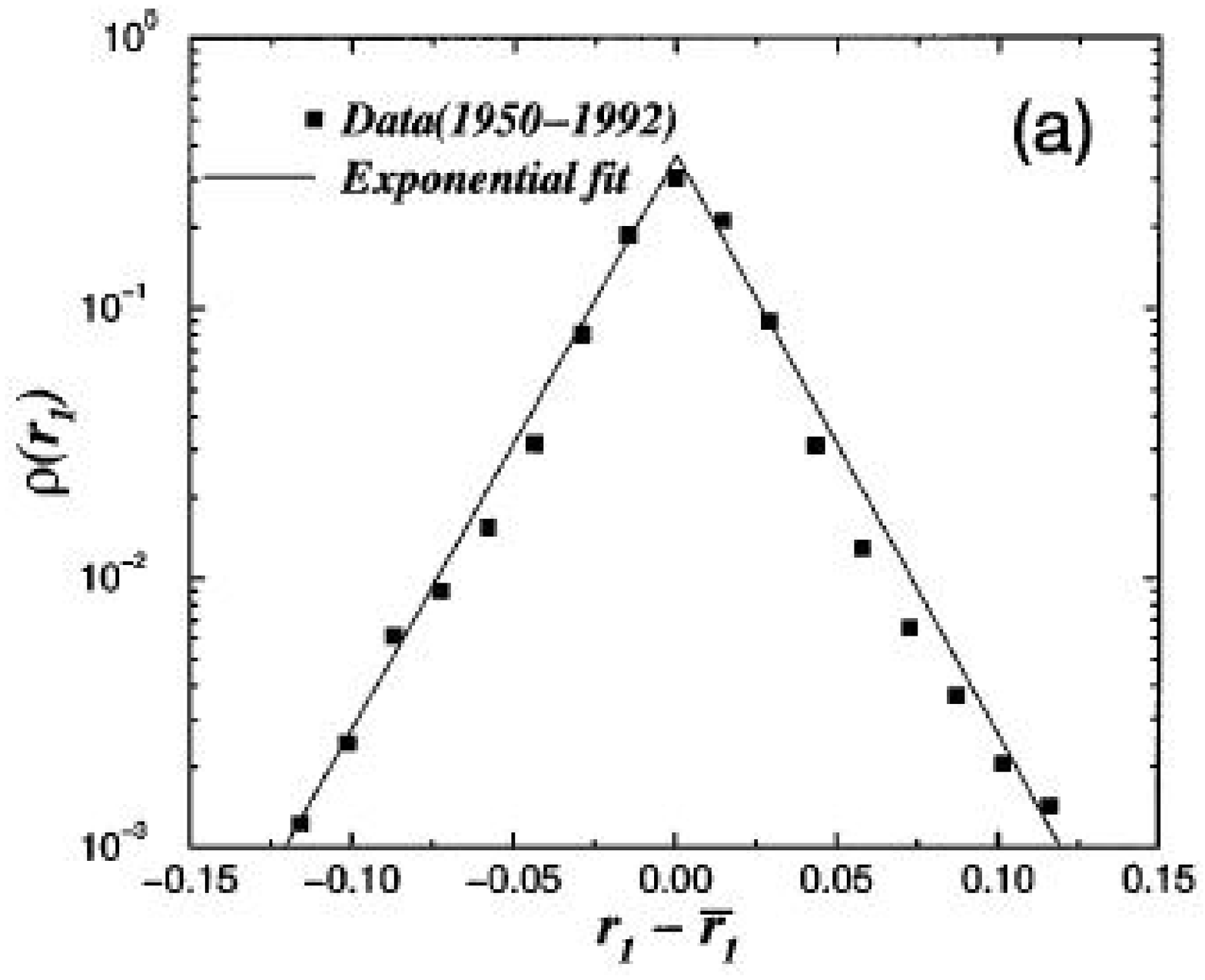}
\includegraphics[width=3in]{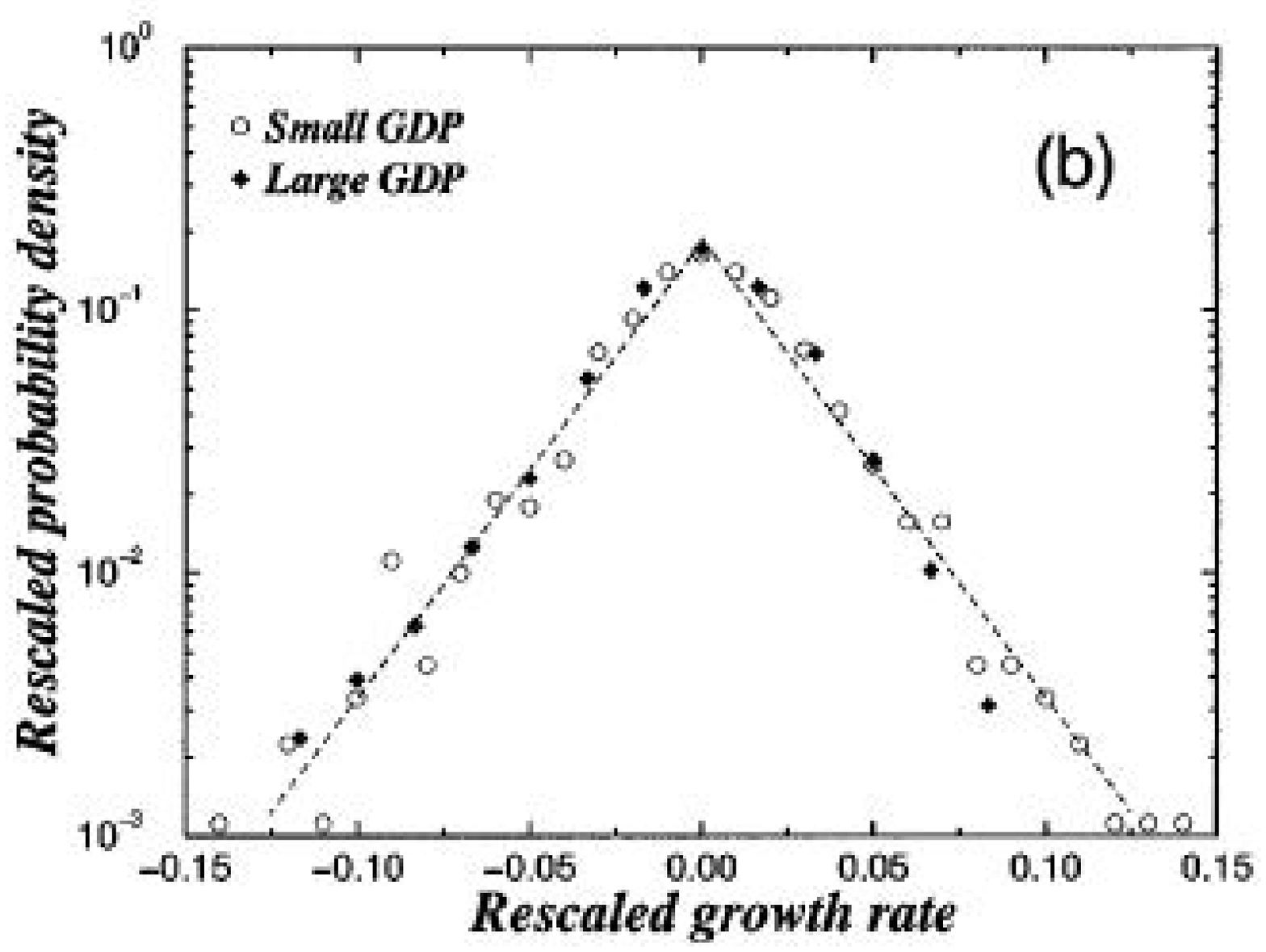}
\caption{Growth rate distribution of firm sizes and GDPs }{extracted from
\cite{firmgrowth, GDP2}, universal Laplace distribution}
\label{figrate}
\end{figure}

At last, we will have conditional distribution function for
different size as $p\left(r|s_{0}\right)$, where $s_{0}$ is the
initial form size. Actually, such analysis is on the first way we
mentioned, keeping eyes on fixed firm, so we get
$p\left(r|s_{i0}\right)$, where $i$ is the label of firm. But
here, a little further we go, the tracks starting at the same size
are combined together. The distribution of growth rate is shown in
figure \ref{figrate} as Laplace distribution,
\begin{equation}
p\left(r|s_{0}\right)=\frac{1}{\sqrt{2}\sigma\left(s_{0}\right)}\exp{\left(-\frac{\sqrt{2}\left|r-\bar{r}\right|}{\sigma\left(s_{0}\right)}\right)}.
\end{equation}

Similar growth rate analysis has also been done for GDP. The gross
growth rate of GDP is defined as
\begin{equation}
p_{i}\left(t\right)=\ln\left(\frac{G_{i}\left(t+1\right)}{G_{t}\left(t\right)}\right).
\end{equation}
But since the long term growth trend of economy, when we want to
analysis the fluctuation information, such endogenous unknown
trend has to been excluded. In \cite{GDP2}, the author suggested
to use a decomposition as below,
\begin{equation}
p_{i}\left(t\right) = \delta_{i} + \phi\left(t\right) +
r_{i}\left(t\right),
\end{equation}
where $\delta_{i}$ is the long term expected endogenous growth
rate, $\phi\left(t\right)$ is a common fluctuation to all
counties, and $r_{i}\left(t\right)$ is the residual which
represents fluctuation, the one we want to investigate. It shows
the same Laplace distribution as shown in figure \ref{figrate}.

\subsubsection{Relation between fluctuation and size}
From experience in Statistical Physics, relation between
fluctuation and size usually gives important information of the
underlying processes\cite{firmgrowth}. Like in idea gas with
independent particles, the magnitude of fluctuation is invert of
the square root of the system size as
\begin{equation}
\sigma\left(N\right)\sim N^{-\frac{1}{2}}.
\end{equation}

\begin{figure}[tbp]
\centering
\includegraphics[width=3.5in]{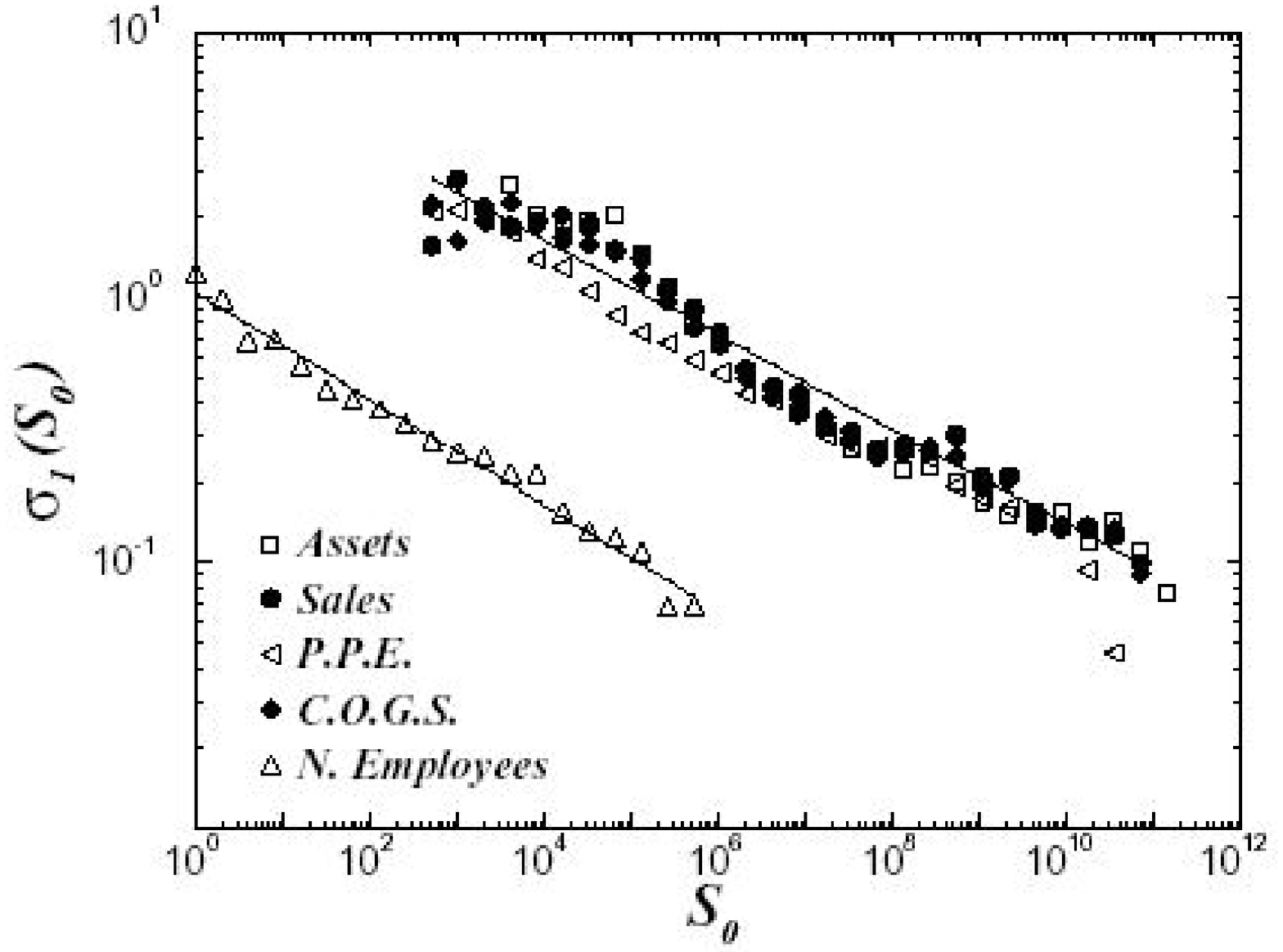}
\includegraphics[width=3.5in]{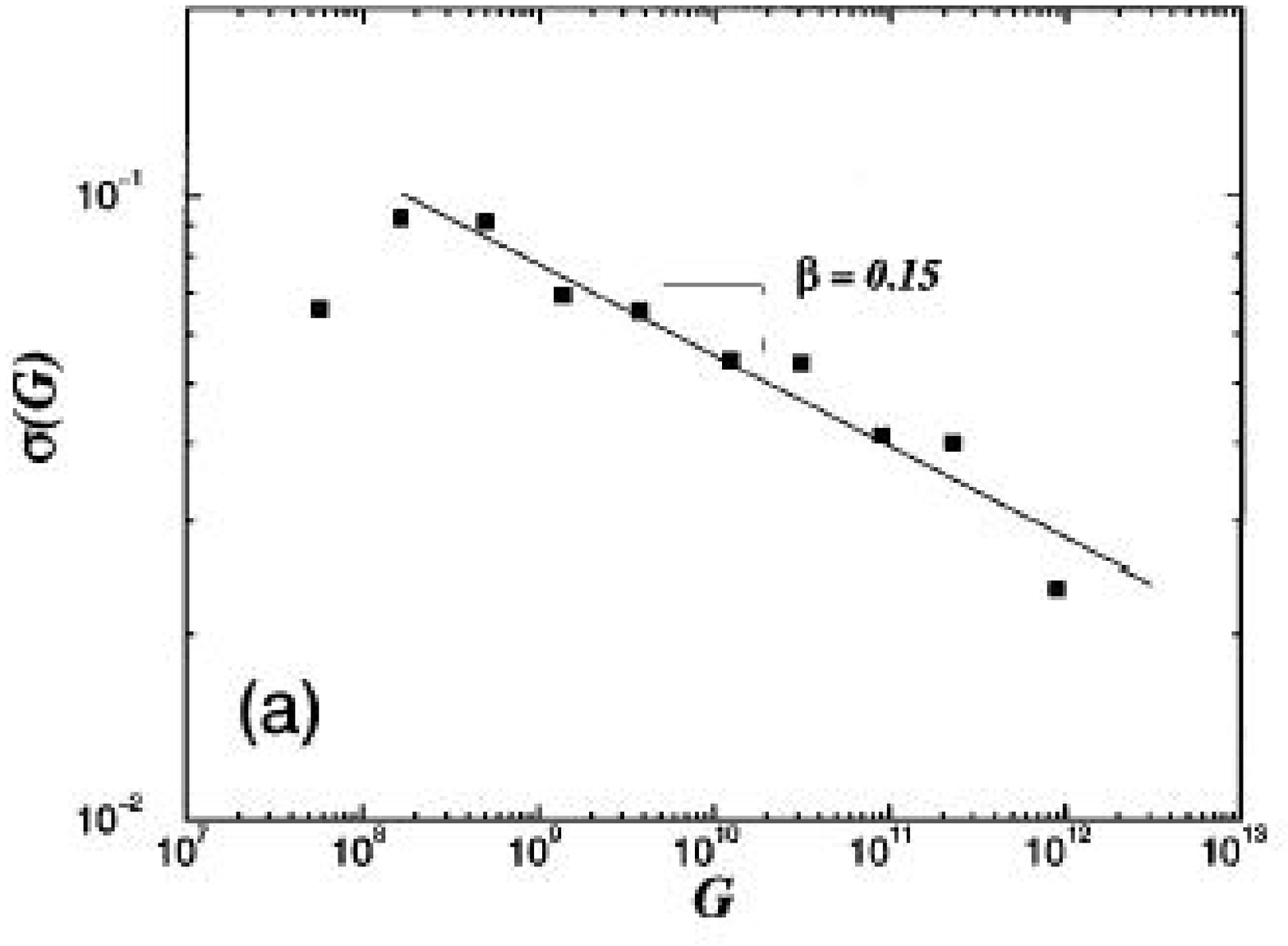}
\caption{Relation between variation and size}{extracted from
\cite{firmgrowth, GDP2}, Power law with similar exponent near $0.15$.}
\label{figfluct}
\end{figure}

Corresponding analysis can be done for growth rate of firm size
and GDP. A power law relation but with exponent other than
$-\frac{1}{2}$ has been revealed by
researchers\cite{firmgrowth,GDP2}. And further more, such
exponents is universal for different measurement of firm size,
independent on time and locations, and the values for firm size
and GDP is very close. Results are shown in figure \ref{figfluct}.
So maybe this implies some common mechanism for firms and GDPs.

\subsection{Complex Networks of economy systems}
Economy is a many-body system including agents as individuals,
firms, countries, goods as produce, production and service, and
subsystems as financial system, manufacturing, agriculture,
service industry. And all of them interact with each other. A
general way developed recently to describe such system is Complex
Networks. In a complex network, every agent is represented by a
vertex and the interaction between any two agents is described by
a link between the two corresponding vertexes. Further more, the
weight of links can be used as the strength of the interaction and
a directed link can be used when the interaction is not
symmetrical.

A recent such development is the web of
trade\cite{tradeweb1,tradeweb2}, in which vertexes are the
countries and links are the inport/outport relation. The basic
structure and efficiency has been analyzed, like high clustering
coefficient, scale-free degree distribution.

Another widely used network of economy system is the interaction
between stock agents. Every stockholder is a vertex in the
network, and the effect from decision of one agent to another is a
directed link from the former vertex to the later. So the network
acts as a whole system to drive the stock price. The geometrical
character of such network will have some important effects on the
dynamical behavior of stock price. Therefor, such investigation
maybe will reveal the interaction pattern between stock agents.

The third proposed works about network of economy systems is the
network analysis of product input/output table. Like the
Predator-Prey Relationships in food web, every product made from
other products or raw materials, and also become input of other
products. So the input/output relationship between products forms
a network. Actually the input/output table analysis in Economics
has the same spirit but in a highly clustered level and asking for
different questions. So, although a database of product relation
is what we need, a clustered group product relation data set will
also be able to be used here as an beginning analysis of basic
structure characters. And further works will require detailed data
on input/output relation of products.

Construction and analysis of such clustered product network is in
progress\cite{klaus}. Characters on degree distribution,
clustering coefficient, weight and weight distribution, average
shortest distance have been gotten, but questions about the
universality of such properties need to be tested on more
networks. The links between products can be regarded as
technology. So a score analysis such as link betweenness will show
the relative importance of different technics, therefor it may
imply some new direction of development of technology. Further
questions about the robustness of such networks can be asked as
how many total product will be lost if one or several
inter-products were in shortage, or when the resource distribution
was changed, or as how many total product will be lost when one or
several link (technics) were dismissed, or inversely, if a new
link was invented how many product will grow in total. Such
investigation will relate traditional questions in economics such
as resource allocation, social welfare, and effect of new
technology with network analysis of product. It can have a
far-reaching effect both on economics and network analysis.

\section{Why is Econophysics?}
We took a review of Econophysics including empirical study and
models on three topics above. Now we try to discuss the question
why is Econophysics? Since dynamics of Stock price is also a topic
of Mathematical financial, what's the difference between this and
Econophysics? If Physics provide insightful tools for this new
field, can Physics also benefit from it?

\subsection{Physics as tools}
First, let's discuss the role of Physics as tools in Econophysics,
the application of concepts, models and method developed in
Physics into Econophysics.

\subsubsection{Physics as analysis method}
There is many-years experience to deal with many-body system and
complex system in Physics. The concepts and technics such as
ensemble statistics, correlation and self-correlation analysis,
have been widely used to reveal the property of economy phenomena.
And the more important thing here is experience in Physics helps
to understand what the properties imply. For instance, a power law
is usually related with critical phenomena in Physics, including
critical point of equilibrium and non-equilibrium phase
transition. And so does a long range order and a high
self-correlation. Also as pointed in \cite{firmgrowth}, relation
between variation and size imply the form of interaction.

Another central analysis method transplanted from Physics is Data
Collapse and Universality. If relation curves from different
systems can be collapsed onto a master curve by scaling, it's very
possible to find some common mechanism from such systems. And if
an empirical or theoretical relation is independent on time
period, some different detail of objects, it's called
universality. When a universal law is found for different systems,
the systems must be equivalent in some ways. So it implies common
mechanism and others can be understood if we know one of them very
well. Therefor, it open a new way to investigate such systems,
especially when some models with similar properties in Physics and
other fields can be used here as a reference model for economy
phenomena.

\subsubsection{Physics as reference model}
Spin model is widely used to describe human decision in stock
market\cite{spin} or other economy activity\cite{division}.
Usually, status of an agent can be one of $\left\{1,0,-1\right\}$,
which is interpreted as buying, waiting and selling, or one of
$\left\{1,-1\right\}$. So the status space of the whole system
with $N$ agents is $\left(S_{1},S_{2},S_{3},\cdots,S_{N}\right)$.
The benefit of every agent is determined by a payoff function
$E\left(\vec{S},\vec{J},IEs\right)$, in which $\vec{J}$ are the
interaction constants of all orders and $IEs$ are the internal
variables as stock price, or external information like environment
and behavior of enterprise. Everyone intend to maximize its own
benefit in a statistical way\cite{spin, division} like
\begin{equation}
\omega_{i}\left(S_{i}\left(t\right)\rightarrow
S_{i}\left(t+1\right)\right)\sim e^{\frac{\Delta E_{i}}{T}},
\end{equation}
in which $T$ is an average evaluation coefficient, which means the
effect on ones decision for a unit benefit. Actually such form of
human decision comes from the ensemble distribution in Statistical
Mechanics. In metropolis simulation of a spin system, the
probability for a spin to transfer its status is overruled by a
similar form. An ensemble distribution here means in statistical
way, in a many body system, although everyone try to stay on
maximum position, but the end status is much like an ensemble
distribution.

Such application gives some reasonable results, although it may be
not totally equivalent with assumption in Economics, where every
agent must stay on its maximum point, not a distribution function.
In Mechanics, the status of physical object is determined by
Newton's equations or minimum action principle, but for a
many-body system in Statistical Mechanics, ensemble distribution
is used instead. Although it's not deduced from first principle,
it works widely. Maybe similar approach can be developed in
Economics.

Ideal gas is another reference model widely used in
Econophysics\cite{money}. In a first order approximation model of
competition and cooperation between firms, or between individuals,
every agent can be regarded as random exchange wealth with each
other, like random exchange energy in ideal gas. So the
equilibrium distribution will take the exponential form. It's
amazing that the central part of personal wealth is actually
exponential form. Further possible model can be generalized random
interaction model, including not random exchange, but also random
increase or decrease process, or extended model with bias exchange
model, like preferential exchange, in which rich one has higher
probability to get richer.

\subsection{Economics as Physics}
In above section, we discussed the role of Physics in
Econophysics. In this section, we ask for the inverse question,
what Physics can benefit from Econophysics, not only as an
insightful tool.

\subsubsection{Economy systems as physical objects}

Frankly, physicists are kinda aggressive, so is Physics. When a
question asking for reason of a phenomenon in a common sense, or
in a fundamental way, according to physicists it's a physical
question. Econophysics is such an example. It choose the special
phenomena from Economics, and ask for the reason, or mechanism in
physical language. Most phenomena concerned in Econophysics
exhibit universality independent on time period, detail of
systems, and even different economical structure of countries. So
such question is likely very much to ask the behavior of a system
with known interactions, or the interaction form of a system with
known behaviors. It's a typical physical question.

Like DFA method proposed by researchers in Statistical Physics
from works in DNA sequence and physiologic signals, new technics
can also be invented from Econophysics. Hopefully, not only
technics, but also concepts and fundamental approach may also be
proposed.

For instance, effect of geometrical property such as dimension and
curvature on dynamical behavior is an important question in
Physics. Actually it's widely studied in Physics including
Relativity, Quantum Physics and especially Phase Transition and
Critical Phenomena. So if geometrical quantities can be defined in
Complex System, and the effect on dynamical process is known, it
will partially predictable just through grasping the geometrical
properties of such systems. For example, in principle, the
make-from relationship between all products is tractable. So the
network can be explicitly constructed, and even part of the
history is known, like the things changed when reformation of
technics happened. So Economics provide some nearly perfect
treasure for Physics. And further more, the special character of
such network will definitely require new quantities or technics to
describe the effect. This will maybe boost the development of
Physics.

\subsubsection{Natural parts of human behavior}

Not all human behaviors are rational or determinant, like
impulsion and inspiration, but some of them, are determined by
environment at least in a statistical way. Personal character
affects human decision. But if all other factors could be
determined by physical way like a dynamical equation, and the
statistical properties of personal characters of the system were
known, it will be easy to predict the behavior of the system. So
the most valuable question left here is that whether we can
describe economy system and human behavior by physical way as far
as possible and leave something unknowable in physical sense. If
it's possible, how to do it. I think, Econophysics is a good try
in such sense.

Economics is a science of human behavior, but it's fortunate that
Economics is not totally a science of human creativity and
inspiration like fine art. This means that some part, even most
part according mathematicians working in Economics, of Economics
can be modelled in an abstract or mathematical form. It's
interesting to point out that it's Physics the most famous
masterpiece applying Mathematics into nature, not any other field
of Applied Mathematics. So it's natural to incorporate Physics
into Economics like to imitate masterpieces.

And through such exploration, it's possible that Physics will be
widely used in social science. This will greatly extend the scope
of Physics, and maybe will help Physics to deal with some hard
topic such as turbulence, or more general complex systems.

\section{Conclusion -- Is Econophysics a subject of Physics?}
At lease, Econophysics provides, invents and develops tools for
analysis of Economy phenomena, and investigation of economy system
generalizes the scope of Physics. But will Econophysics effect
concepts and thought in Physics? It depends on the future.
However, we are sure that both Economics and Physics can benefit
from such exploration. Therefor, as a researcher in physics in the
new century, or a potential economist, should we learn from each
other?

\section{Acknowledgement}
Thanks is given to Fukang Fang and Zhanru Yang for their
simulating discussion, and to graduate students 2002 in System
Science Department for their warm discussion and good questions.
The author Wu want to give thanks to Qian Feng for her
encouragement and understanding. This work is partially supported
by National Natural Science Foundation of China under the Grant
No. 70371072 and No. 70371073.

\end{document}